\documentclass[aip,jvst,reprint]{revtex4-1}
\usepackage{graphicx}
\usepackage{ulem}
\usepackage{amssymb}
\usepackage{amsmath}
\usepackage{wasysym}
\usepackage{relsize}
\usepackage{siunitx}
\begin{document}

\newcommand{\be}{\begin{equation}}
\newcommand{\ee}{\end{equation}}
\newcommand{\bea}{\begin{eqnarray}}
\newcommand{\eea}{\end{eqnarray}}
\newcommand{\Tbar}{{\bar{T}}}
\newcommand{\En}{{\cal E}}
\newcommand{\calK}{{\cal K}}
\newcommand{\U}{{\cal U}}
\newcommand{\GC}{{\cal \tt G}}
\newcommand{\Lop}{{\cal L}}
\newcommand{\DB}[1]{\marginpar{\footnotesize DB: #1}}
\newcommand{\q}{\vec{q}}
\newcommand{\kt}{\tilde{k}}
\newcommand{\Lopn}{\tilde{\Lop}}
\newcommand{\noi}{\noindent}
\newcommand{\ovn}{\bar{n}}
\newcommand{\ovx}{\bar{x}}
\newcommand{\ovE}{\bar{E}}
\newcommand{\ovV}{\bar{V}}
\newcommand{\ovU}{\bar{U}}
\newcommand{\ovJ}{\bar{J}}
\newcommand{\calE}{{\cal E}}
\newcommand{\ovphi}{\bar{\phi}}
\newcommand{\zt}{\tilde{z}}
\newcommand{\ttl}{\tilde{\theta}}
\newcommand{\nuv}{\rm v}
\newcommand{\ds}{\Delta s}
\newcommand{\fn}{{\small {\rm  FN}}}
\newcommand{\cc}{{\cal C}}
\newcommand{\cd}{{\cal D}}
\newcommand{\tth}{\tilde{\theta}}
\newcommand{\cb}{{\cal B}}
\newcommand{\cg}{{\cal G}}
\newcommand\norm[1]{\left\lVert#1\right\rVert}

\title{A hybrid approach to modelling large area field emitters}

\author{Debabrata Biswas}

\affiliation{
Bhabha Atomic Research Centre,
Mumbai 400 085, INDIA}
\affiliation{Homi Bhabha National Institute, Mumbai 400 094, INDIA}

\begin{abstract}
  Large area field electron emitters, typically consisting of several thousands of nanotips,
  pose a major challenge since numerical modelling requires enormous computational resources. We propose
  a hybrid approach where the local electrostatic field enhancement parameters of an individual emitter is
  determined numerically while electrostatic shielding
  and anode-proximity effects are incorporated using recent analytical advances. The hybrid model is
  tested numerically on an ordered arrangement of emitters and then applied to recent
  experimental results on randomly distributed gold nanocones.
  Using the current-voltage data
  of two samples with vastly different emitter densities but having similar nanocone sizes, we show that an
  appropriate modelling of the emitter-apex together with the analytical results on shielding and anode-proximity
  effects, leads to consistent results for the apex radius of curvature.
  In both cases, the $\text{I-V}$ data is approximately reproduced for $R_a \simeq 9$nm.
  Importantly, it is found that anode-proximity plays a significant role 
  in counter-balancing electrostatic shielding and ignoring this effect results in the requirement of
  a much smaller value of $R_a$.
  \end{abstract}

\maketitle

\section{Introduction}
\label{sec:intro}

Large area field emitters (LAFE) are promising candidates as cathodes in
fast-switching vacuum-electronic devices and have found use in
miniature x-ray sources, space applications,
vacuum gauges and even domestic lighting\cite{teo,dams2012,wilfert2012,li2015,hong2018,sheshin2019,ohkawa2019}.
While ordered LAFE have long been investigated \cite{spindt76,spindt91,whaley2009,helfenstein},
a random distribution of emitters is also of interest and occurs for instance when dealing
with carbon nanotubes \cite{read_bowring}.

\begin{figure}[hbt]
  \begin{center}
\hskip -0.8cm
\hspace*{-2.9cm}\includegraphics[scale=0.15,angle=0]{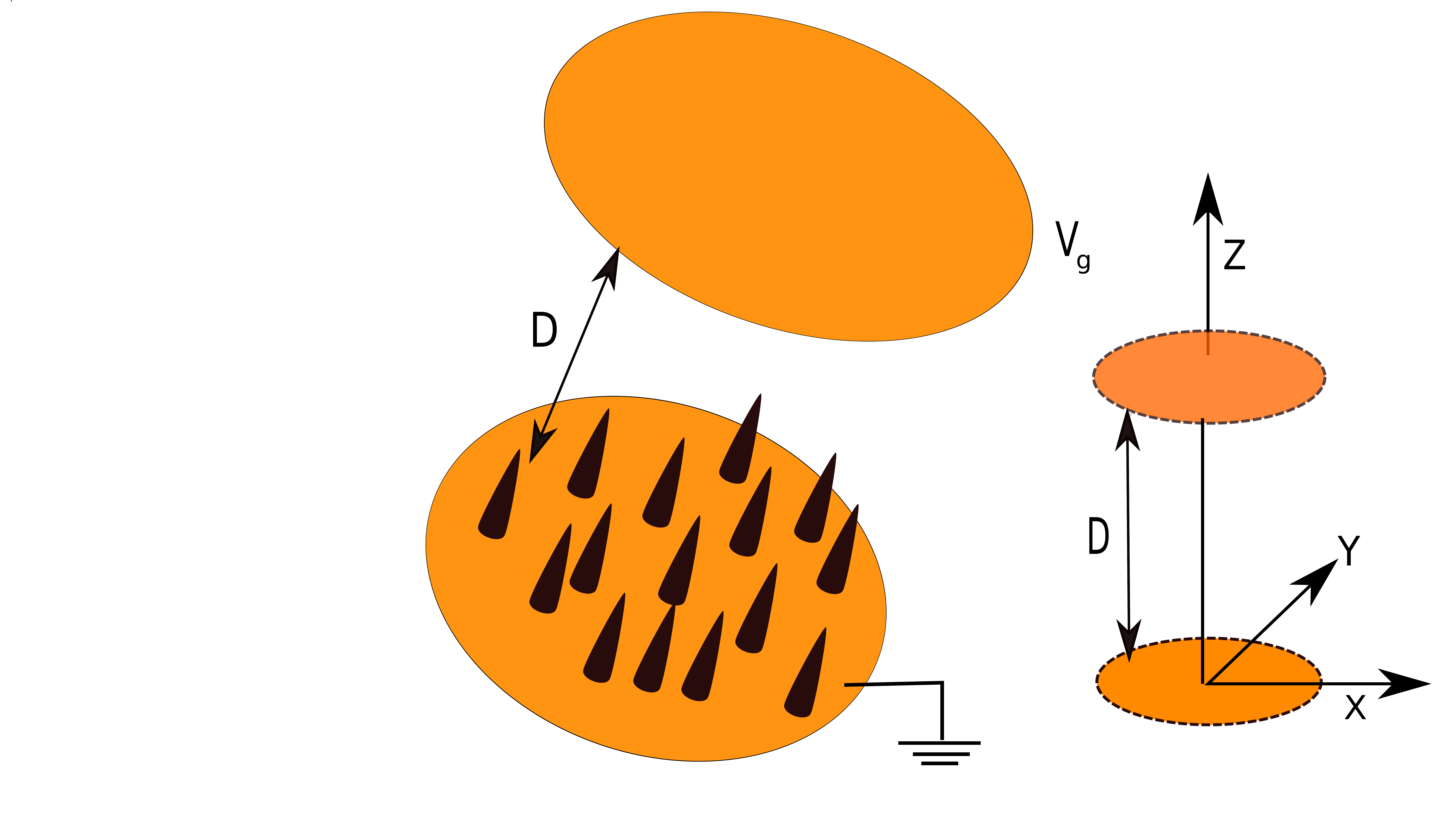}
\vskip -0.25 cm
\caption{Schematic of a LAFE with the anode at a distance $D$ and
  at a potential $V = V_g$ while the cathode plate is grounded.
   The actual LAFE may contain tens of thousands of emitters
   and can be ordered instead of being randomly distributed.
   Also shown alongside is the co-ordinate system.  
}
\vskip -0.5 cm
\label{fig:LAFE}
\end{center}
\end{figure}

Recently, there have been studies involving metallic micro-structured cathodes with randomly distributed
nanocones (see Fig.~\ref{fig:LAFE}) grown using irradiated polymer templates \cite{bieker2018}.
An advantage of such a technique is the near-identical
height of the emitting structures and a good degree of control over the shape. Since the nanocones are
metallic in nature, it also offers a chance to interpret the results using standard field emission
theory which assumes the free-electron model\cite{FN,MG,forbes_deane,jensen_ency,kyritsakis2015,
  db_parabolic,db_curvature}.
A direct application of the theory however requires
knowledge about the local field at each of the several thousand emitter apex. In a uniform random distribution of
emitters, the local field can vary enormously from emitter to emitter so that an evaluation of the field
emission current is prone to large errors unless the distribution of
the apex field enhancement factors, accounting for both
electrostatic shielding and anode-proximity effects, is taken into consideration \cite{dist_known,db_rudra,levine95}.

A random large area field emitter thus poses a major challenge since the number of emitters that can be modelled
numerically (using for example, finite element software such as COMSOL) is severely limited
by computational resource requirements. A recent study\cite{bieker2019} used 25 emitters, while it is unlikely
that more than a hundred random emitters can be simulated using
reasonable present-day resources \cite{deassiss2020}.
A finite-sized ordered collection of emitters can be equally challenging since each emitter may need
to be accounted for while determining the local field.
Thus, a hybrid approach seems necessary where analytical inputs can be combined
with numerical simulation to handle tens of thousands of emitters or an even larger collection.

In the following sections, we shall first outline this hybrid approach (section~\ref{sec:hybrid})
and then apply it to the specific experiment involving gold nanocones (section~\ref{sec:results}).
Finally, we shall discuss the methodology used and draw conclusions from the analysis.

\section{The hybrid approach to modelling field emission from LAFE}
\label{sec:hybrid}

The hybrid model, as the name suggests, is a combination of the two approaches. The {\it first} is the
purely numerical approach, particularly finite element modelling\cite{assis2019}, which has been extremely successful
in determining the local field at the apex of (i) a single isolated emitter (ii) a
small collection of a emitters, typically less than 100 (iii) as well an infinite number of emitters
placed in an ordered manner (for example on a 2-dimensional square lattice) which can be simulated using
suitable boundary conditions on the  computational domain boundaries.
The {\it second} approach is analytical, where recent applications\cite{db_fef,db_anodeprox,rr_db_2019,db_rr_2020}
of the nonlinear\cite{db2016} line charge
model\cite{mesa,pogo2009,harris15,harris16} (LCM) have yielded several
inputs that can be combined in a modular fashion to approximate
fairly well the local field at the apex of thousands of individual emitters comprising a LAFE.
In this second approach, a  general expression for the local field at the apex of
the $i^{th}$ emitter reads as

\be
E_a^{(i)} \simeq E_0 \frac{2h/R_a}{\alpha_1 \ln(4h/R_a) - \alpha_2 - \sum \text{contributions}}
\ee

\noi
where $\alpha_1$ and $\alpha_2$ depend on the shape of the 
an isolated individual emitter with the anode far away. 
The sum over {\it contributions} take into account all the factors that affect the local field
at the apex due to extraneous conditions such as the proximity to the anode,
the direct shielding effect of other emitters and the effect of other emitters mediated
through the anode. The values of $\alpha_1$ and $\alpha_2$ are constants that
are unknown in general\cite{alp1alp2,db2016,db_fef} except in special shaped protrusions
where the exact solution to the Laplace equation is known. They must thus be determined
for an arbitrary shape. Fortunately
however, the {\it contributions} can be expressed approximately in terms of purely
geometric quantities such as the anode-cathode gap $D$, the height $h$, and the distance on
the cathode plane between pairs of emitters, all of which are known for a given LAFE.

The hybrid model thus seeks to use the {\it first} approach to determine
$\alpha_1$ and $\alpha_2$ and then use the analytical expressions for {\it contributions}
to determine the local field at the apex of each emitter in a LAFE. Once the local fields
at the apex are known, contemporary field emission theory can be used to compute the emission current.
The following 2 sub-sections
give details of how the local field can be determined followed by a test of its accuracy.

\subsection{An isolated emitter in a diode configuration}
\label{sec:isolated}

The first step in the present hybrid approach is an expression for the apex field enhancement factor $\gamma_a$
of an isolated emitter placed in a planar diode configuration with the anode far away\cite{edgcombe2002,forbes2003,db_fef}.
For any {\it isolated} axially symmetric emitter of height $h$ and apex radius of curvature $R_a$ placed normal to
the cathode in a parallel plate diode configuration, the nonlinear line charge model\cite{db2016} has been used to express the 
apex field enhancement factor as \cite{db_fef}

\vskip -0.5cm
\be
\gamma_a \simeq \frac{2h/R_a}{\alpha_1 \ln(4h/R_a) - \alpha_2}   \label{eq:gam_basic}
\ee

\noi
where $\gamma_a = E_a/E_0$. Here $E_a$ denotes the electric field at the emitter apex while $E_0$ is
the macroscopic field far away from the emitter. In a planar diode configuration where the anode-cathode
distance is $D$ and the potential difference is $V_g$, the macroscopic field $E_0 = V_g/D$. Eq.~(\ref{eq:gam_basic})
is a good approximation when $h/R_a$ is sufficiently large and the anode is far away ($D >> h$).
The quantities $\alpha_1$ and
$\alpha_2$ are constants for a particular emitter-shape. For instance\cite{jensen_ency}, in case of a hemiellipsoid
$\alpha_1 = 1$ while $\alpha_2 = 2$. For another shape such as a nanocone with a rounded apex for which
$\alpha_{1,2}$ cannot be analytically determined,
a numerical approach can be used. Thus, $\gamma_a$ can be computed
numerically (for instance using COMSOL) for various values of $h/R_a$ in the regime of interest
and then a plot of $(2h/R_a)/\gamma_a$ vs $\ln(4h/R_a)$ can be used to extract the values of
the parameters $\alpha_1$ and
$\alpha_2$ that characterize individual isolated emitters\cite{shreya_db_2019}.
Since the points are expected to be on a straight line,
the quantities $\alpha_1$ and $\alpha_2$ are respectively the slope and intercept. An expression for the
apex field enhancement factor for a single isolated emitter with the `anode-at-infinity' can thus be
determined using nominal computational resources \cite{db_fef,shreya_db_2019}.
This will be illustrated in more detail in Section \ref{sec:test}.

\subsection{A non-isolated emitter in a diode configuration}

The presence of the anode close to the emitter, the presence of other emitters with the anode
far away or the presence of other emitters with the anode in close proximity, are some of the
situations which affect the local field at the apex of an emitter. They make the emitter {\it non-isolated}
and under the influence of these extraneous conditions.

\subsubsection{The presence of anode}

The presence of the anode (or gate) is the logical next step in building the hybrid model.
We shall therefore ignore the presence of other emitters and focus on how this factor alone
affects the field at the apex.

It is well known that the presence of the anode in close proximity to the emitter apex results
in an increase in the apex field enhancement factor \cite{wang,smith}. For a hemiellipsoidal emitter, it has been
shown that \cite{db_anodeprox}

\be
\gamma_a(D) \simeq \frac{2h/R_a}{\alpha_1 \ln(4h/R_a) - \alpha_2 - \alpha_A}  \label{eq:gam_anode}
\ee

\noi
where $\alpha_A$ is the {\it contribution} due to anode proximity on
an otherwise isolated emitter. It depends on $h$ and $D$ and can be expressed as\cite{db_anodeprox}

\be
\begin{split}
\alpha_A = & \sum_{n=1}^\infty \Bigg[ \frac{(2nD - h)}{h} \ln\Big(\frac{2nD - h + L}{2nD - h - L}\Big) \\
  & - \frac{(2nD + h)}{h} \ln\Big(\frac{2nD + h + L}{2nD + h - L}\Big) \Bigg] \label{eq:alpA}
\end{split}
\ee

\noi
with $L = h - R_a/2$.
Eq.~(\ref{eq:alpA}) strictly holds for hemiellipsoidal emitters, but can be used
approximately for other emitter shapes\cite{db_anodeprox}. It has been used
in conjunction with Eq.~(\ref{eq:gam_anode}) for  (i) a paraboloid
(ii) a paraboloid on cone and (iii) a hemiellipsoid on a cylindrical post. It was
found that once the quantities $\alpha_1$ and $\alpha_2$ are determined for each of
these shapes, Eq.~(\ref{eq:gam_anode}) can be used to determine the field enhancement
factor at various anode-cathode distance with good accuracy.

A single emitter with the anode in close proximity can thus be modelled using the
hybrid approach as outlined above. This has been confirmed numerically even for
non-ellipsoidal emitter shapes\cite{db_anodeprox} and the predictions are within acceptable limits
(generally less than $5\%$ error) even when $D$ is only slightly larger than $h$.

\subsubsection{A collection of emitters with the anode far away}

We next consider a large area field emitter with the anode far away. This gives
rise to the shielding effect which leads to a reduction in local field
at the apex of an individual emitter.

\begin{figure}[hbt]
  \begin{center}
    \vskip -1.5cm
\hspace*{-0.60cm}\includegraphics[scale=0.275,angle=0]{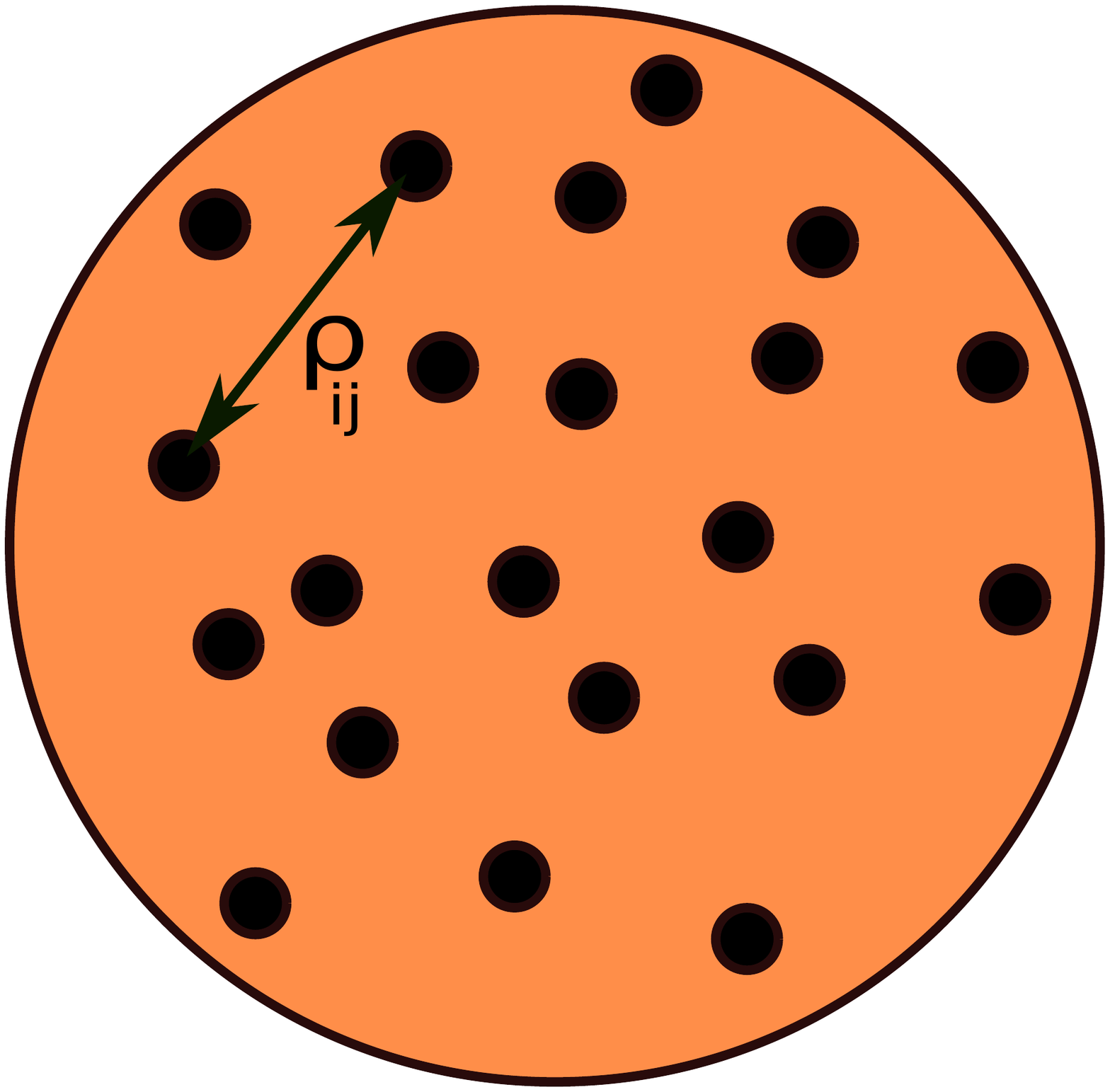}
\vskip -1.5 cm
\caption{A cross-sectional view of a typical LAFE. The nano-emitters are marked in black.
  The distance between $i^{th}$ and $j^{th}$ emitter is denoted by
  $\rho_{ij} = \sqrt{(x_i - x_j)^2 + (y_i - y_j)^2}$ where ($x_i,y_i$) is co-ordinate
  of the centre of the $i^{th}$ emitter. The hybrid model holds equally for an
  ordered placement of emitters.
}
\label{fig:cross_section}
\end{center}
\end{figure}

Consider a LAFE consisting of $N$ axially symmetric single-emitters, each of height $h$ and apex
radius of curvature $R_a$, in a parallel plate diode
configuration (see Fig.~\ref{fig:LAFE}) with the cathode
at $z = 0$ and the anode far away ($D >> h$).  Let the distance between the
$i^{th}$ and $j^{th}$ emitter be denoted by $\rho_{ij}$ as shown in Fig.~\ref{fig:cross_section}.
The shielding effect on the apex field enhancement factor of the $i^{th}$ emitter
can be expressed as \cite{db_rudra}

\be
\gamma_a^{(i)} \simeq \frac{2h/R_a}{\alpha_1 \ln(4h/R_a) - \alpha_2 - \alpha_{S_i}}  \label{eq:gam_shield}
\ee

\noi
where $\alpha_{S_i}$ is the {\it contribution} due to the shielding effect of other emitters when
the anode is far away. It can be expressed as \cite{db_rudra}

\be
\begin{split}
  \alpha_{S_i} \simeq & \sum_{j\neq i}^{N}  \Bigg[ \frac{1}{h}\sqrt{\rho_{ij}^2 + (h - L)^2} - \frac{1}{h}\sqrt{\rho_{ij}^2 + (h + L)^2} \\
    & + \ln\Bigg(\frac{\sqrt{\rho_{ij}^2 + (h + L)^2} + h + L}{\sqrt{\rho_{ij}^2 + (h - L)^2} + h - L}\Bigg) \Bigg] \label{eq:alpS}
\end{split}
\ee

\noi
and $L = h - R_a/2$. 

Note that Eq.~(\ref{eq:gam_shield}) is independent of the arrangement of $N$ emitters and
hold equally well for ordered as well as a random distribution.
The accuracy of Eq.~(\ref{eq:gam_shield}) in predicting the local field at the apex
has recently been tested \cite{rr_db_2019}
and found to be satisfactory so long as the emitters are not
too close to each other ($\rho_{ij} > h/2$).

\subsubsection{A collection of emitters with the anode in close proximity}

Finally, we shall study the most general case while dealing with a
large area field emitter. Consider therefore a collection of $N$ emitters
placed normally on the cathode plate with the anode close by such that its
effect cannot be ignored. 
Both, the direct effect of the anode and neighbouring emitters as well as
the indirect effect of neighbouring emitters (through a succession of images of opposite induced
charge polarity due to the presence of the anode and cathode planes) can be incorporated to express
the apex field enhancement factor of the $i^{\text{th}}$ emitter as \cite{db_rr_2020}

\be
\gamma_a^{(i)} \simeq \frac{2h/R_a}{\alpha_1 \ln\big(4h/R_a\big) - \alpha_2 - \alpha_A +  \alpha_{S_i} - \alpha_{{SA}_i}} \label{eq:gamSA}
\ee

\noi
where $\alpha_{{SA}_i}$ is the {\it contribution} due to the indirect effect of neighbouring
emitters mediated through the
anode and cathode planes. It helps in countering the shielding effect when the anode is sufficiently close
and can be expressed as\cite{db_rr_2020}

\be
\begin{split}
  \alpha_{{SA}_i} \simeq & \sum_{n=1}^\infty \sum_{j\neq i}^{N}  \Bigg[
    \frac{\cd_{mm}}{h} - \frac{\cd_{mp}}{h} - \frac{\cd_{pm}}{h} + \frac{\cd_{pp}}{h} \\
    & + \frac{2nD - h}{h} \ln\Big(\frac{\cd_{mp} + 2nD - h + L}{\cd_{mm} + 2nD - h - L}\Big) \\
    & - \frac{2nD + h}{h} \ln\Big(\frac{\cd_{pp} + 2nD + h + L}{\cd_{pm} + 2nD + h - L}\Big) \Bigg] \label{eq:alpSA}
\end{split}
\ee

\noi
with

\bea
\cd_{mm} & = & \sqrt{\rho_{ij}^2 + (2nD - h - L)^2} \nonumber \\
\cd_{mp} & = & \sqrt{\rho_{ij}^2 + (2nD - h + L)^2} \nonumber \\
\cd_{pm} & = & \sqrt{\rho_{ij}^2 + (2nD + h - L)^2} \nonumber \\
\cd_{pp} & = & \sqrt{\rho_{ij}^2 + (2nD + h + L)^2}. \nonumber
\eea

\noi
In the above $\rho_{ij} = \sqrt{x_{ij}^2 + y_{ij}^2}$ is the distance between the $i^{\text{th}}$ and $j^{\text{th}}$
emitter on the cathode plane ($\text{XY}$) and $L = h - R_a/2$.

Note that unlike $\alpha_A$, the expressions for $\alpha_{S_i}$ and $\alpha_{{SA}_i}$ are
approximate even for a hemiellipsoid emitter and can be used\cite{rr_db_2019,db_rr_2020}
if a pair of emitters is not
closer than $h/2$. The approximation gets better as the pairwise distance increases.
While it is difficult to test the result numerically for a large collection of
emitters, Eq.~(\ref{eq:gamSA}) has been verified for an infinite LAFE of hemiellipsoidal emitters
on a 2-dimensional square lattice. It was found to give
excellent results when the lattice constant is $c = 1.5h$. The results are
reasonable when $c = h$ while for $c < h/2$, the error in prediction increases significantly.

\subsection{Testing the hybrid approach}
\label{sec:test}

\begin{figure}[hbt]
  \begin{center}
    \vskip -3.0cm
    \hspace*{-0.1cm}\includegraphics[scale=0.35,angle=0]{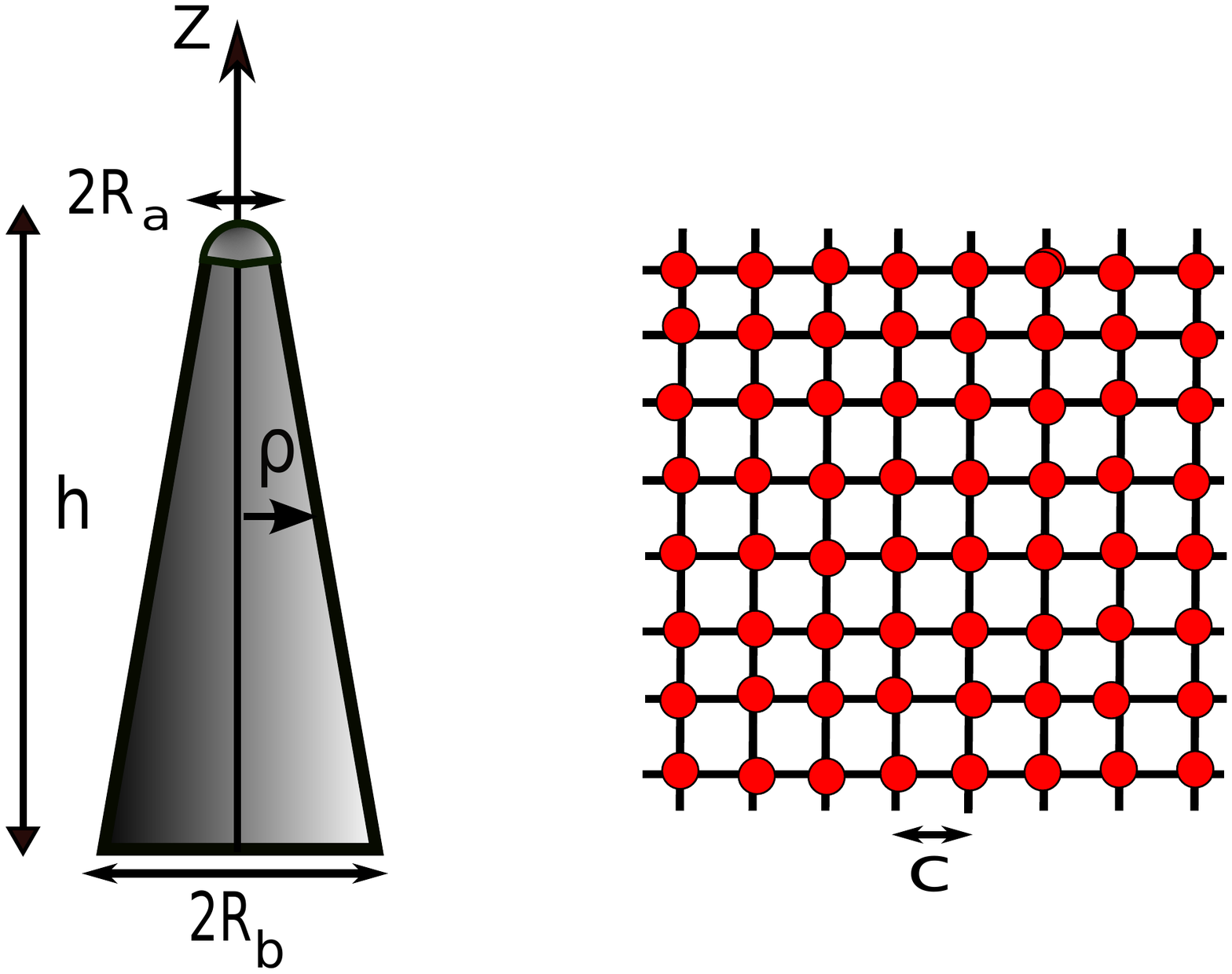}
\vskip -3.0 cm
\caption{An individual emitter is modelled as a circular cone of base radius $R_b$, total height $h$ with a hemi-spherical
  endcap of radius $R_a$ on top. The radial distance $\rho$
  measures the transverse distance from the $Z$-axis.
  On the right, the red circles mark the location of the nanocones on the 2-dimensional
  square lattice having lattice constant $c$. 
}
\label{fig:cone}
\end{center}
\end{figure}

The expressions for $\alpha_A$, $\alpha_S$ and $\alpha_{SA}$ are approximate for non-hemiellipsoidal
emitter shapes and hence the efficacy of the hybrid approach needs to be tested.
Consider thus a LAFE consisting of circular-cones, each having
a hemispherical endcap (see Fig.~\ref{fig:cone}). In order to test the hybrid approach,
we shall consider an ordered LAFE of infinite extent
on a square lattice having lattice constant $c$ as in Ref. [\onlinecite{db_rr_2020}] except
that the hemiellipsoid is replaced by circular-cones with a hemispherical endcap.

The rounded conical emitters are considered identical and each has
a total height $h = 24\mu$m, base radius $R_b = 1.75\mu$m and apex radius of
curvature $R_a = 0.01\mu$m.
The emitter shape and the dimensions have been chosen in order to be close
to the experimental situation  \cite{bieker2018} to be modelled eventually in the following
section. In the same spirit, let the anode-cathode
distance $D = 48 \mu$m.   The value of $c$ determines the the density of emitters
and can be varied to determine its effect on the accuracy of the hybrid model.

Note that the isolated rounded nanocone can be modelled in several ways. We have chosen here a nanocone
with a hemispherical end-cap at the apex. Other possibilities include a hemi-ellipsoidal or
parabolic end-cap and while the value of $\gamma_a$ does depend on the choice of the end-cap,
a test of the hybrid approach is largely independent of its specific nature. The total height
of the cone and endcap is thus $24\mu$m and the hemispherical part has a height $0.01\mu$m.

\begin{figure}[hbt]
  \begin{center}
    \vskip -1.0cm
\hspace*{-0.50cm}\includegraphics[scale=0.34,angle=0]{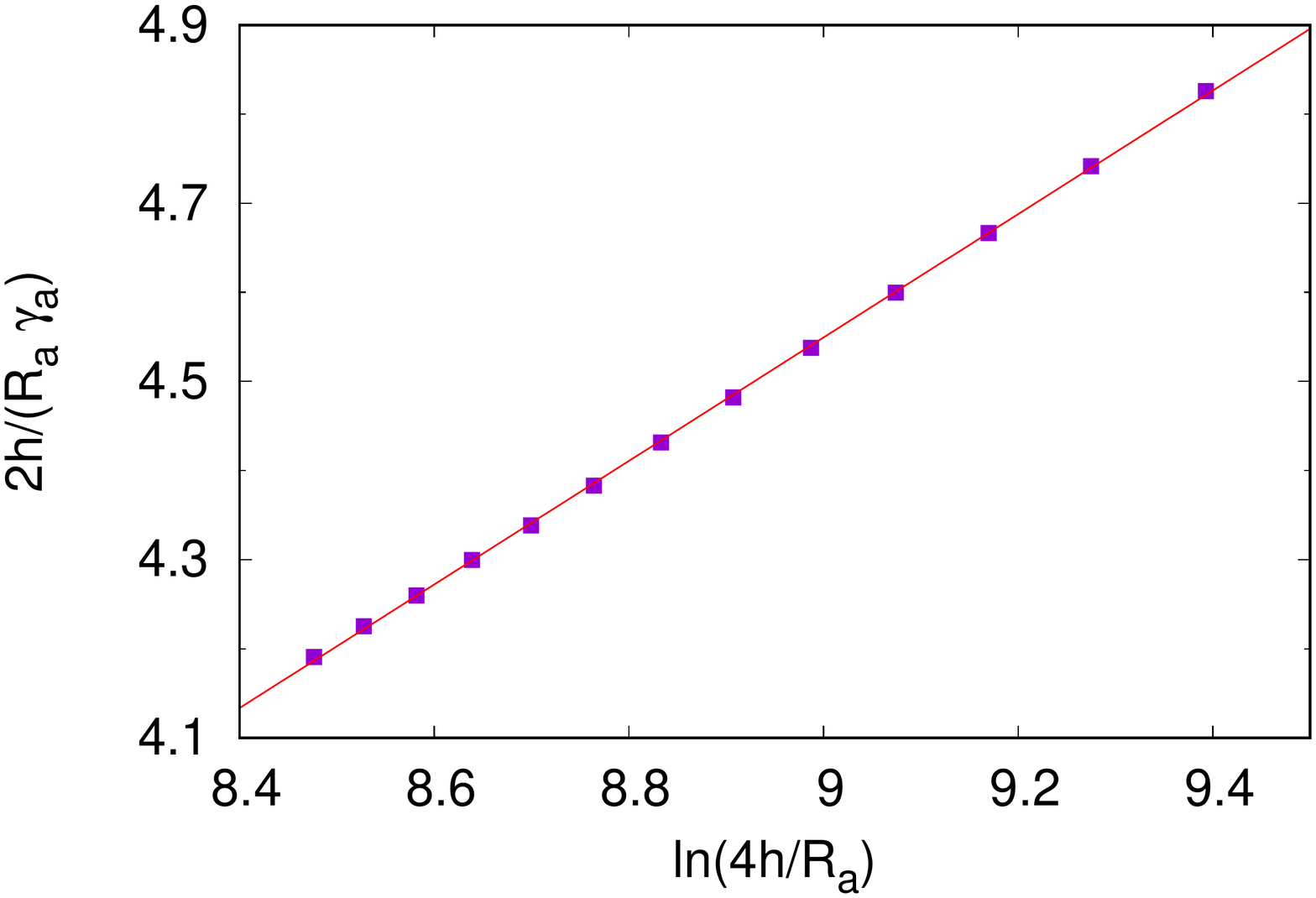}
\vskip -0.5 cm
\caption{The nanocone height $h$ is kept fixed at $24\mu$m while $R_a$ is varied. In each case, the apex
  field enhancement factor $\gamma_a$ is determined using COMSOL v5.4. In order to simulate an isolated
  emitter, the domain boundaries are kept sufficiently far away with appropriate boundary conditions.
  The straight line fit (solid line) is used to determine $\alpha_1$ and $\alpha_2$.
}
\vskip -0.5cm
\label{fig:gam_isolated}
\end{center}
\end{figure}

The prescription outlined in section \ref{sec:isolated}
requires us to estimate $\alpha_1$ and $\alpha_2$ by fitting a
straight line (see Ref. [\onlinecite{shreya_db_2019}] for other examples)
to a plot of $(2h/R_a)/\gamma_a$ vs $\ln(4h/R_a)$. This can be achieved by
varying $R_a$ around the value chosen ($0.01\mu$m) while keeping $h$ fixed.
In each case, $\gamma_a$ is obtained using COMSOL v5.4. The emitter is modeled
as a perfect electrical conductor which is at ground potential along with the cathode plane.
The boundary condition at the anode placed at $z = D$
can be Dirichlet ($V = V_g$, where $V_g$ is the anode potential) or
Neumann ($\partial V/\partial z = \epsilon_0 E_0$).
For simulating an isolated emitter with the anode-at-infinity, the Neumann boundary
condition is preferred\cite{assis2019,rr_db_2019} since $D$ need be no more than $5h$. The transverse
boundaries are similarly kept at $X,Y = \pm 5h/2$ and the Neumann boundary
condition $\partial V/\partial(x,y) = 0$ is imposed. For each simulation,
convergence in $\gamma_a$ is verified using finer meshing.

Fig.~\ref{fig:gam_isolated}
shows the data (solid squares) corresponding to $R_a \in [7,20]\text{nm}$ along with the
best fit straight line corresponding to $\alpha_1 = 0.693$ and $\alpha_2 = 1.69$.
Recall that the first step in the hybrid approach requires us to determine
the values of $\alpha_1 = 0.693$ and $\alpha_2$ numerically and that is now achieved.

The next step requires us to evaluate $\alpha_A$,
$\alpha_S$ and $\alpha_{SA}$ using the values of $c$ and the anode-cathode gap $D$.
For a square lattice, $\rho_{ij} = c\sqrt{(m_i - m_j)^2 + (n_i - n_j)^2}$ where
$\vec{\rho_i} = m_i \hat{x} + n_i \hat{y}$, $\vec{\rho_j} = m_j \hat{x} + n_j \hat{y}$
are the position vectors of the $i^{th}$ and $j^{th}$ cone on the $XY$ plane.
Finally, the values of $\alpha_1$, $\alpha_2$, $\alpha_A$,
$\alpha_{S_i}$ and $\alpha_{SA_i}$
can be used to evaluate $\gamma_a^{(i)}$ using Eq.~(\ref{eq:gamSA}).

In order to simulate an infinite square lattice using COMSOL,
the `zero surface charge density' boundary
condition is imposed on the transverse boundaries  of the computational
domain (i.e at $X,Y = \pm c/2$) having a nanocone at its centre.
The results obtained for the apex field enhancement factor can then be
compared with the values obtained using Eq.~(\ref{eq:gamSA})
to get an estimate of the error in hybrid modelling for different values of
$c$.

\begin{figure}[hbt]
  \begin{center}
    \vskip -0.85cm
\hspace*{-0.50cm}\includegraphics[scale=0.34,angle=0]{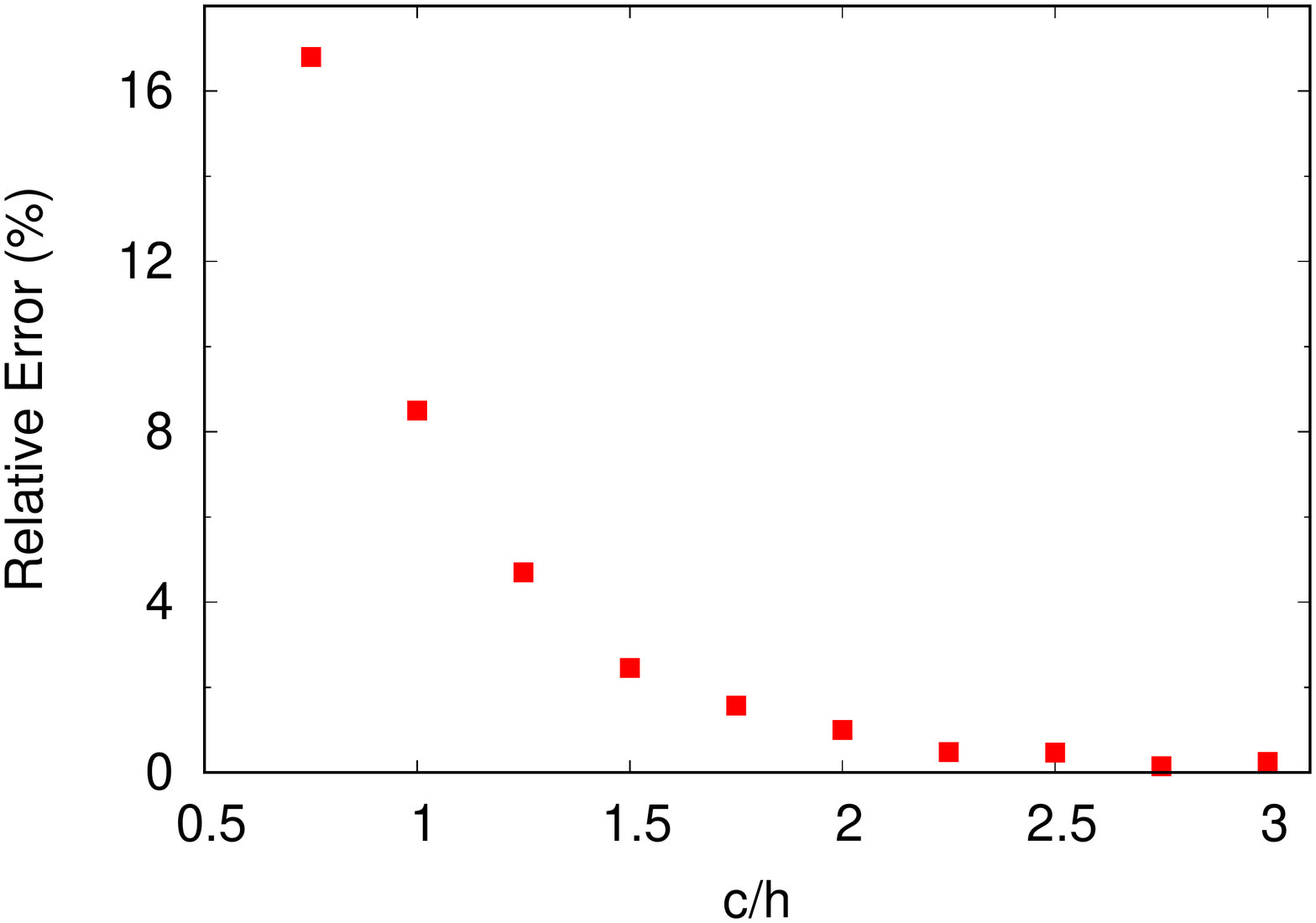}
\vskip -0.5 cm
\caption{The relative error (Eq.~\ref{eq:error}) in predicting the apex field enhancement factor using the
  hybrid approach for different values of the lattice constant. The error gets larger
  as the distance between emitters decreases.
  }
\label{fig:gam_error}
\vskip -2.0cm
\end{center}
\end{figure}

Fig.~\ref{fig:gam_error} shows the relative error for different values of the lattice constant $c$.
The relative error is defined as

\be
\text{Relative Error}(\%) = \frac{ |\gamma_a^{\text{comsol}} - \gamma_a^{\text{hybrid}}|}{\gamma_a^{\text{comsol}}} \times 100  \label{eq:error}
\ee

\noi
where $\gamma_a^{\text{comsol}}$ is the value determined using COMSOL
while $\gamma_a^{\text{hybrid}}$ is computed using Eq.~(\ref{eq:gamSA}). The error is small at larger spacings.

\begin{figure}[hbt]
  \begin{center}
    \vskip -0.85cm
\hspace*{-1.0cm}\includegraphics[scale=0.34,angle=0]{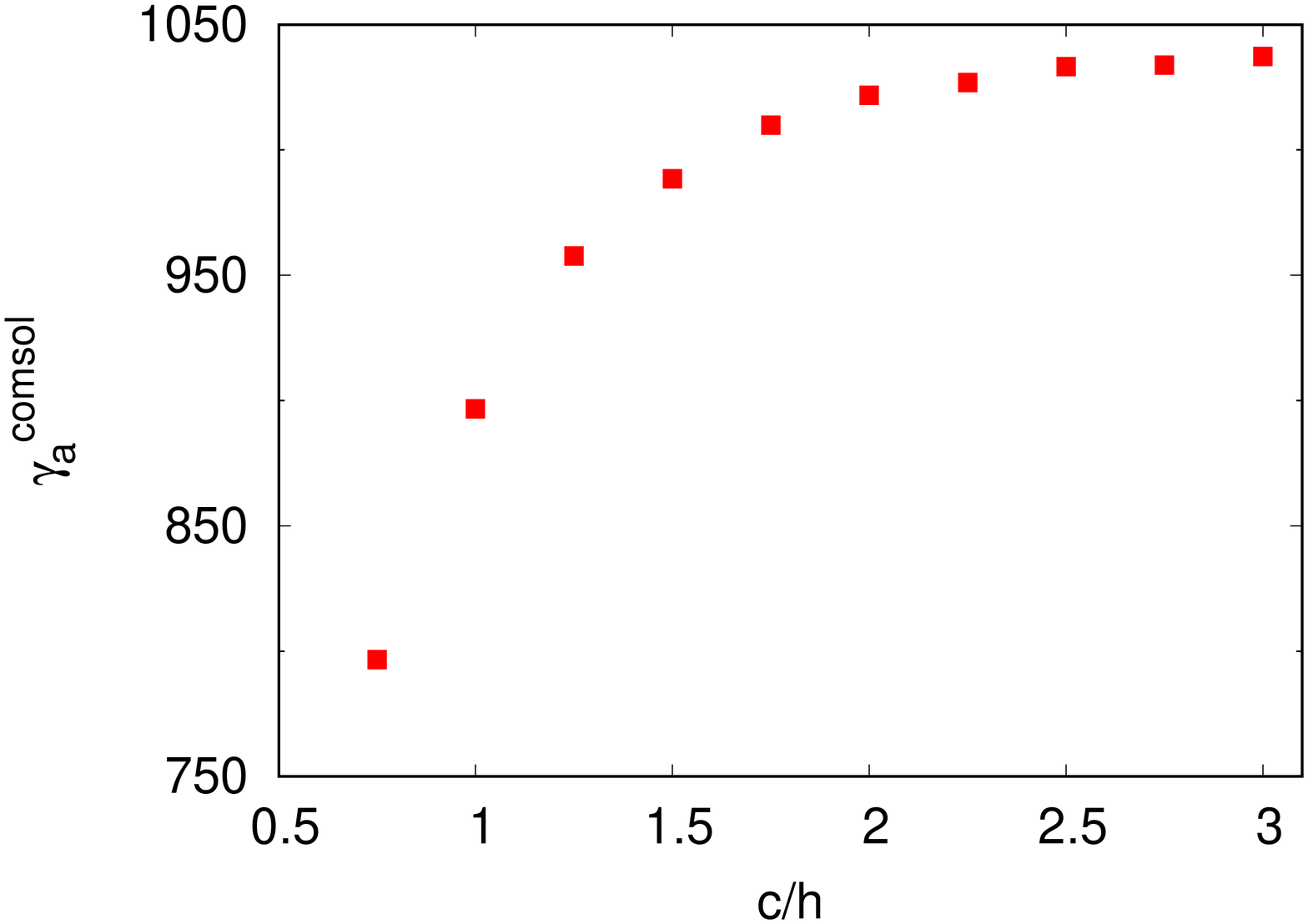}
\vskip -0.5 cm
\caption{The change in apex field enhancement factor determined using COMSOL as the lattice constant $c$ is
  varied.  The anode-cathode gap $D = 48\mu$m.
  }
\label{fig:gam_vary}
\end{center}
\end{figure}

Fig.~\ref{fig:gam_vary} shows the variation of $\gamma_a$  with the lattice constant $c$.
The enhancement factor falls sharply for $c < h$ when shielding finally overcomes the
direct and indirect field enhancing effect of the anode.
Interestingly, the error is small when $\gamma_a$ is
large. The implications of this observation for randomly placed large area field emitters  is enormous since
the ones that do contribute significantly to the current are expected to be relatively isolated, having
large field enhancement and hence a smaller error in the hybrid model prediction.

\section{Application to field emission from randomly distributed nanocones}
\label{sec:results}

In the previous section, the hybrid model was introduced and found to work for an ordered
large area field emitter with reasonable accuracy.
For $c = h$, the relative error was found to be about 8\%, reducing rapidly\cite{error1} to about 2\% for $c = 1.5h$ .
For randomly distributed emitters, the results for field emission current are expected to be better than
ordered LAFEs since the few emitters that do contribute have large field enhancement factors (being relatively
isolated) compared to the bulk of emitters that suffer shielding due to close proximity to other
emitters. Thus, the hybrid model can be used to make useful predictions about field emission
current for randomly distributed emitters as well.

In the following, we shall apply the hybrid model
to a recent experiment \cite{bieker2018} using randomly distributed identical conical
field emitters where the height ($h$) of
an individual emitter and the mean density of emitters is known. The only unknown
parameter is the apex radius of curvature $R_a$ for which only an upper bound
is reported\cite{bieker2018}.

\subsection{The experiment and the choice of endcap in the model}

The experiment reported in Ref.~[\onlinecite{bieker2018}] involves
field emission from a random distribution of gold nanocones on a circular patch.
The nanocones were grown using ion-track etched
polymer templates. The heavy ion bombardment and a subsequent asymmetric wet-etching of the polymer foil
leads to a uniform random distribution of conical pores. The template is then fixed onto a
metallized glass substrate, structured to give it a circular shape and
the pores in the template filled using electro-deposition to obtain
gold nanocones. Finally, the polymer template is removed via chemical etching. Further details of
the process and SEM images of the samples produced can be found in Ref.~[\onlinecite{bieker2018}].

Three samples were reported, each having nanocones of height $h = 24\mu$m but having
different densities. Sample A had a density ($\text{cones/cm}^2$) of $6 \times 10^4$, sample B had $4\times 10^5$
while sample C had $1\times 10^6$. The base diameter of the cone was $3.0\mu$m for sample A,  $3.6\mu$m for sample B and
$3.75\mu$m for sample C. Only the upper bound of the tip diameter ($2R_a$) was reported. They were
respectively $500$nm for sample A and $300$nm for samples B and C. The area of the {\it circular} LAFE cathode
was reported to be $4.9 \text{mm}^2$.

In the field emission experiments using these samples, the anode-grid was placed at a
distance $D = 50\mu$m from the cathode plane and the $\text{I-V}$ data was recorded in
each case. Sample B reported the lowest on-set voltage and reported a maximum current $I_{max} = 142.2\mu$m at
a voltage $V_{max} = 339$V. Sample C fared next and the low emitter-density of sample A led to a
much higher on-set voltage.

Due to the similarities of sample B and C, we shall focus on these to see whether the hybrid model
gives consistent results. The uncertainty in apex radius $R_a$ implies that this must be used as a
free parameter. Our aim is to see whether the $\text{I-V}$ characteristics can be approximately
reproduced for samples B and C for nearly the same value of $R_a$.

Note that apart from the apex radius of curvature, the exact shape of the end-cap is also
unknown. A locally parabolic end-cap follows the generalized cosine law of local
surface field variation
$E(\rho,z) = E_a \cos(\tilde{\theta})$ where $\cos(\tilde{\theta}) = (z/h)/\sqrt{(z/h)^2 + (\rho/R_a)^2}$
in the vicinity of the apex as established in Ref. [\onlinecite{db_ultram,cosine,db_anodeprox,db_rr_2020}].
Here $(\rho,z)$ are points on the surface of the end-cap of an axially
symmetric emitter with its centre at the origin,
while $E(\rho,z)$ is the magnitude of the (normal) field. Since most smooth end-caps
shapes are locally parabolic (i.e. of the form $z \simeq h - \rho^2/(2R_a^2)$ where $\rho = \sqrt{x^2 + y^2}$)
the exact shape does not matter so long as they have identical $h$ and $R_a$. A hemispherical
end-cap on the other hand, does not follow the generalized cosine law. Rather, the local field falls
off slower than the generalized cosine law as we move away from the apex. Thus, a nanocone with a
hemispherical endcap should emit larger current compared to a hemi-ellipsoidal endcap with the same
apex radius of curvature $R_a$. Since the upper bound of $R_a$ as reported in Ref.~[\onlinecite{bieker2018}]
is quite large, we shall choose the hemispherical end-cap in order to obtain the largest possible
value of $R_a$ that can mimic the experimental $\text{I-V}$ curve. Note that the test for the
hybrid model lies in obtaining the $\text{I-V}$ curve for samples B and C using nearly the same
value of $R_a$ for a given end-cap.

As a note of caution, it is worth mentioning that the end-cap need not be smooth
in reality and there might be micro-protrusions on its
surface. Thus the end-cap might have a much larger radius of curvature but the presence
of tiny protrusions may enhance the local field as expected from the Schottky Conjecture \cite{schott23,stern}
or its recent corrected variant \cite{db_schottky}. Such a possibility cannot be ruled out and we acknowledge
that an alternate model based on multiplicative effect may be constructed to yield
a larger value of $R_a$. We shall however restrict ourselves here to the case of smooth end-caps
and explore whether the hybrid model gives consistent results.

\subsection{Variation of field on hemispherical end-cap}

The hybrid model can be used to determine the apex field enhancement factor $\gamma_a$ of individual nanocones
in either sample using Eq.~(\ref{eq:gamSA}) as illustrated in section \ref{sec:test}. In order to compute the
field emission current however, we need to know how
the local field behaves away from the apex on the surface of each emitter. 
On a hemispherical endcap, the generalized cosine law does not apply
and the decay in local field away from the apex is much slower. For the hybrid model to
be applied seamlessly to nanocones with hemispherical endcaps having $R_a$ as a parameter,
it is important that an alternate scaled variation of the electric field exists with $\rho/R_a$.
We shall look into this aspect here.

\begin{figure}[hbt]
  \begin{center}
    \vskip -0.55cm
\hspace*{-.50cm}\includegraphics[scale=0.35,angle=0]{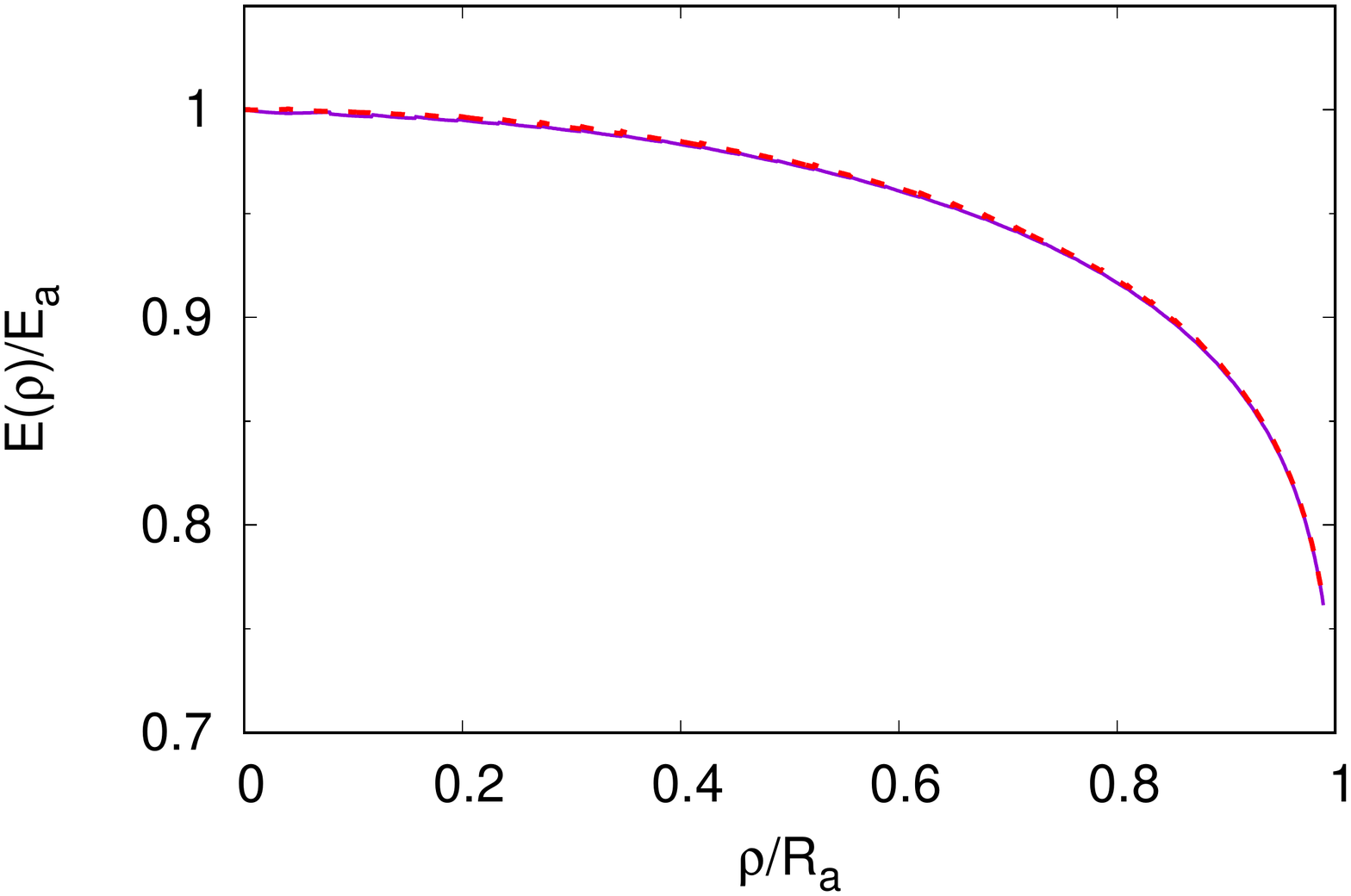}
\vskip -0.75 cm
\caption{The change in field enhancement factor (determined using COMSOL) with $\rho/R_a$ along the surface of
  the hemispherical endcap for 2 different apex radius of curvature. The dashed curve is for $R_a = 15$nm
  while the solid curve is for $R_a = 20$nm.
  }
\label{fig:vary}
\end{center}
\end{figure}

Fig~\ref{fig:vary} shows a plot of the variation of $E(\rho)/E_a$ with $\rho/R_a$, on the surface of a
hemispherical end-cap of radius $R_a = 15$nm and 20nm. Clearly, the scaled variation
is identical for the 2 cases. This has been established
for several values of $R_a$ in the range [5,25]nm and for different lattice constants $c$,
keeping the height $h$ and anode-cathode distance $D$ invariant.

Note that the scaled variation on the hemispherical endcap (as in Fig~\ref{fig:vary})
depends weakly on the cone angle. However, for a given cone-angle or total height, it is
nearly universal with respect to the apex radius of curvature.

Thus, the local field on a hemispherical endcap follows a universal scaled variation which
can be used to determine the net field emission current from a random LAFE.

\subsection{Results for the random LAFE}

We are now in a position to determine the current from a random LAFE using the hybrid model.
The first step towards this is a realization of a uniform random emitter distribution
on the circular patch of area $4.9 \text{mm}^2$ using a standard uniform
random number generator. The points ($x_i,y_i$) thus generated, correspond to the co-ordinates of the
circular nanocone centre. Further, from the experimental data
(see the SEM image in Fig.~3 of Ref. [\onlinecite{bieker2018}]), it is evident that some
of the nanocones either do not grow fully or break in the process of dissolving the
template. Since, the height of these  stubs  is small, they can be ignored
altogether in the simulation as they have negligible shielding effect on other
nanocones. We have assumed that for  samples B and C, 90\% of the emitters are of height
$h = 24\mu$m and the rest can be neglected. Since both samples have equal area (4.9 $\text{mm}^2$),
the number of random emitter-positions simulated
is thus 17640 for sample B and 44100 for sample C. The average spacing in both cases is thus less than
the height of the nanocones. However, as mentioned earlier, only those emitters that are
relatively isolated from the others, are expected to
contribute to the current since they have a higher
value of $\gamma_a$. Since the hybrid model predicts the higher $\gamma_a$ values more accurately,
and underestimates the smaller $\gamma_a$ values, the error in net emission current is expected to
be small.
Note that the number of emitters in either sample is much beyond the scope of a purely numerical
approach due to resource limitations.

Having simulated the emitter positions, the next step is to determine the apex field
enhancement factor of each emitter using the hybrid model (Eq.~(\ref{eq:gamSA})).
This is then used in conjunction with the variation of the local field
on the hemispherical endcap (see Fig.~\ref{fig:vary}), to determine the
current from each emitter by integrating over the hemispherical endcap
\cite{FN,MG,forbes_deane,jensen_ency,kyritsakis2015,db_curvature}

\be
I  =  \int_0^{R_a} J(E(\rho,z)) \sqrt{1 + \left(\frac{dz}{d\rho}\right)^2}~ 2\pi \rho d\rho.
\ee

\noi
With $z$ as the symmetry axis, $\rho$ and $z$ are related on the endcap as $z = \sqrt{R_a^2 - \rho^2}$ so that

\be
I = 2\pi R_a \int_0^{R_{a}} J(\rho) \frac{\rho}{\sqrt{R_a^2 -  \rho^2}}~d\rho
\ee

\noi
is the the net emission current from a single emitter. Note that an appropriate choice of the
current density must be made depending on
whether curvature corrections are to be included. In the present case, since the 
radius of curvature is less than $1\mu$m, the curvature corrected current density \cite{kyritsakis2015,db_curvature,db_gated}
at the point $(\rho,z)$ on the end-cap is used:

\be
J(\rho) =  \frac{1}{\mathlarger{\tilde{t}_F^2(\rho)}} \frac{A_\fn}{W} E^2(\rho) \exp\left(-\frac{B_\fn \tilde{v}_{{\small F}}(\rho) W^{3/2}}{E(\rho)}\right). \label{eq:FNC0}
\ee

\noi
In the above,

\bea
\tilde{v}_{{\small F}} & = & v_{\small F} + {\cal X}_{F}~w_F~,~\tilde{t}_{\small F} =  t_{\small F}  +   {\cal X}_{F}~\psi_F \\
v_{\small F} & = & 1 - f_0 + \frac{1}{6} f_0 \ln f_0 \\
t_{\small F} & = & 1 + \frac{f_0}{9} - \frac{1}{18} f_0 \ln f_0,~{\cal X}_{F}  =  \frac{W}{E(\rho) R_a} \\
w_F & = & \frac{4}{5} - \frac{7}{40} f_0 - \frac{1}{200} f_0 \ln f_0 \\
\psi_F & = &   \frac{4}{3} - \frac{1}{500} f_0 - \frac{1}{30} f_0 \ln f_0 \\  
\cb & = & \frac{B_\fn W^{3/2}}{E_a}~,~f_0  \simeq  1.439965 \frac{E(\rho)}{W^2} \label{eq:last}
\eea

\noi
where  $A_\fn~\simeq~1.541434~{\rm \mu A~eV~V}^{-2}$,
$B_\fn~\simeq 6.830890~{\rm eV}^{-3/2}~{\rm V~nm}^{-1}$ are the conventional FN constants and
$W = 4.8$eV is the work function for gold\cite{bieker2018,bieker2019}.
The current from all the emitters are thus computed and added to determine the net
field emission current for a particular value of $R_a$ from the sample.

\begin{figure}[hbt]
  \begin{center}
    \vskip -0.55cm
\hspace*{-.50cm}\includegraphics[scale=0.35,angle=0]{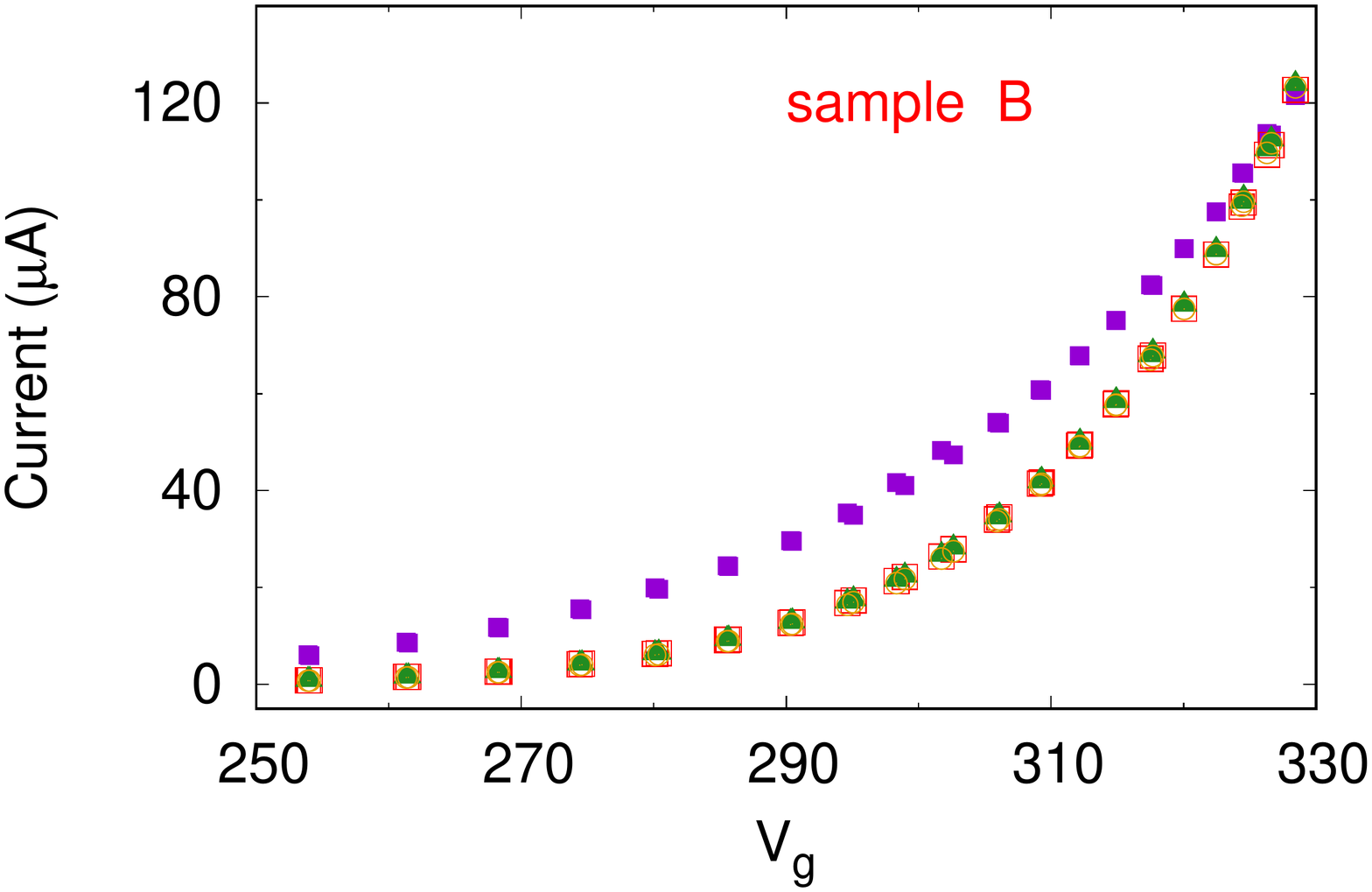}
\vskip -0.75 cm
\caption{The current from sample B. The solid squares are the experimental values
  while the others are the result of computation (solid triangle, unfilled square, and circle)
  using the hybrid model for 3 different
  realizations of the 17640 uniformly distributed emitter positions. The value of $R_a$ used is 8.8nm in all cases.}
\label{fig:sampleB}
\end{center}
\end{figure}

\begin{figure}[hbt]
  \begin{center}
    \vskip -0.75cm
\hspace*{-.70cm}\includegraphics[scale=0.35,angle=0]{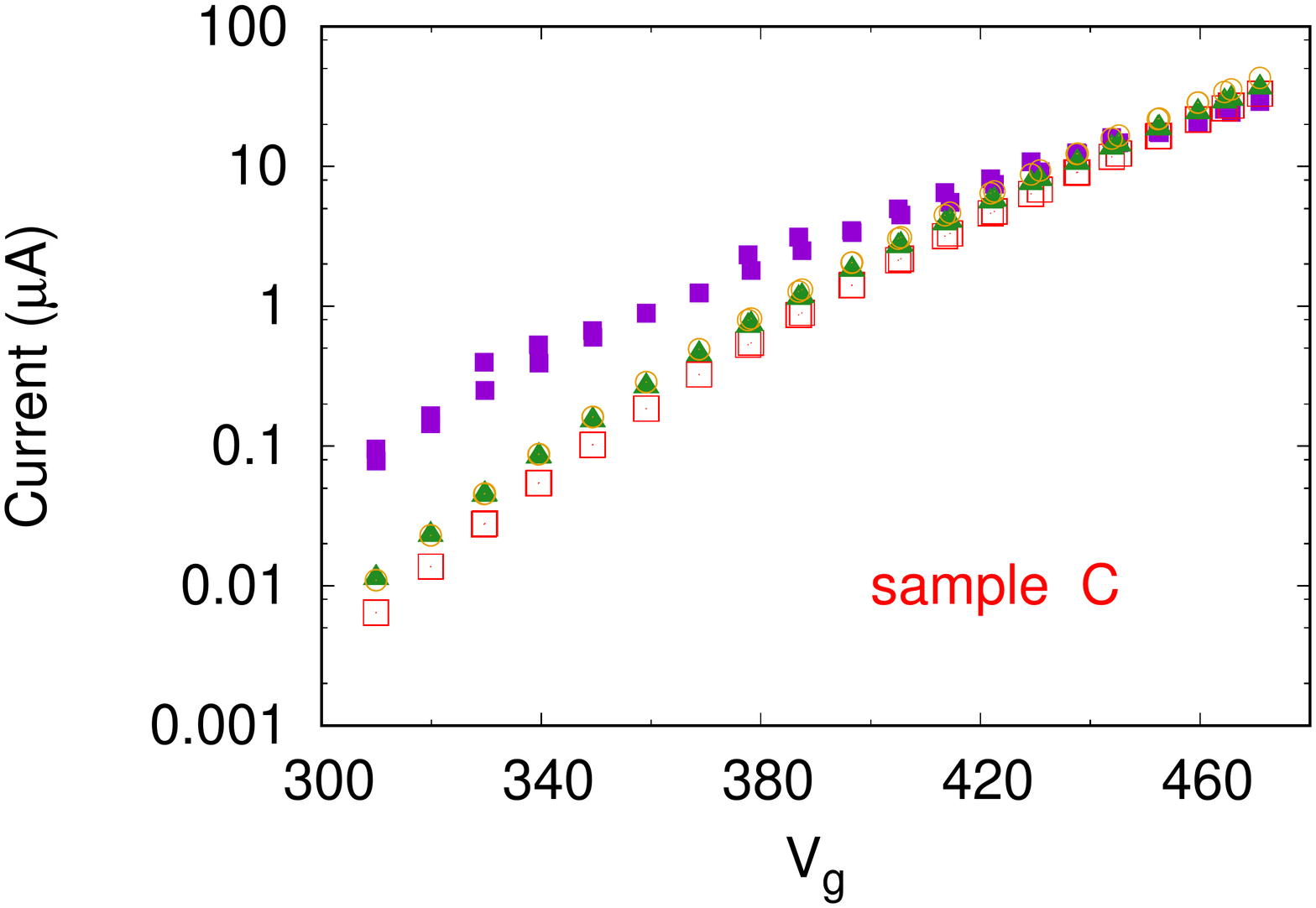}
\vskip -0.85 cm
\caption{The current from sample C. The solid squares are the experimental values
  while the others are the result of computation (denoted by solid triangle, unfilled square, and circle)
  using the hybrid model for 3 different realizations of the 44100 randomly distributed emitter positions.
  The value of $R_a$ used is 9.1nm in all cases.}
\label{fig:sampleC}
\end{center}
\end{figure}

\begin{figure}[hbt]
  \begin{center}
\hspace*{-.40cm}\includegraphics[scale=0.45,angle=0]{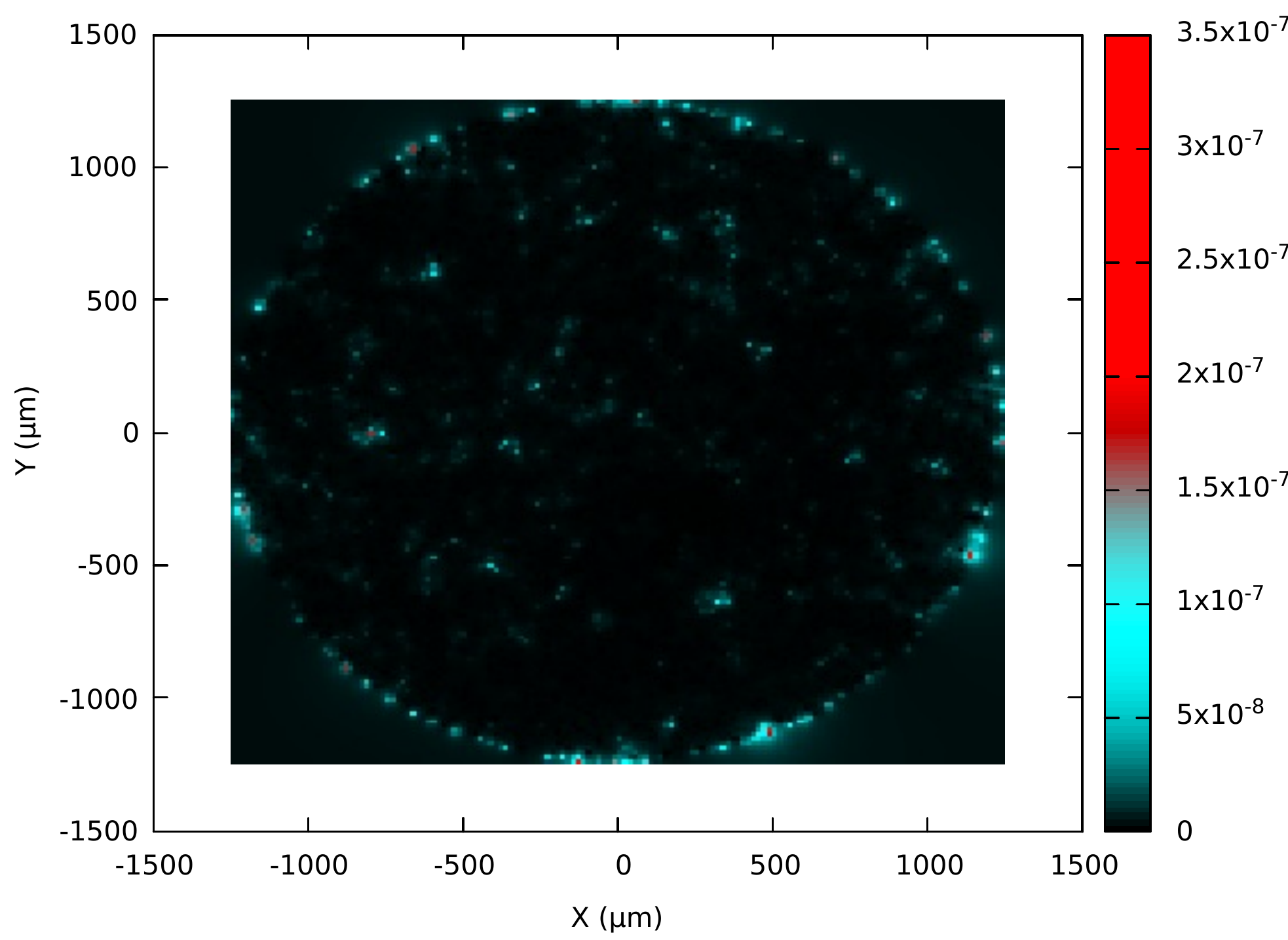}
\caption{A typical current emission map of sample B cathode for a particular realization of
  the random distribution. Emission is visible from the interior region in addition to
  the predominant boundary contribution. Note that the cathode is circular in shape.
 }
\label{fig:map_B}
\end{center}
\end{figure}

Fig.~\ref{fig:sampleB} shows the experimental plot of the field emission current from sample B
together with the current obtained from the hybrid model using three realizations of
emitter positions. Note that the model as well as the field emission theory
being used have several approximations. In fact, the curvature
correction as in Eq.~(\ref{eq:FNC0}) is expected to work best at higher voltages
and larger apex radius of curvature. Since a single parameter ($R_a$) has been varied
to approximate the experimental $\text{I-V}$ plot, a perfect fit is unlikely and
un-physical for a theoretical model that is approximate. In keeping with the
expectations from Eq.~(\ref{eq:FNC0}), we have shown the results of the hybrid model in Fig.~\ref{fig:sampleB}
for the value of $R_a$ that best approximates the current at the highest voltages.
It may be noted that the current from the LAFE is largely independent of the three random realizations
and each of the cases gives identical results across the range of voltage considered.

\begin{figure}[hbt]
  \begin{center}
    \vskip -0.50cm
\hspace*{-.40cm}\includegraphics[scale=0.45,angle=0]{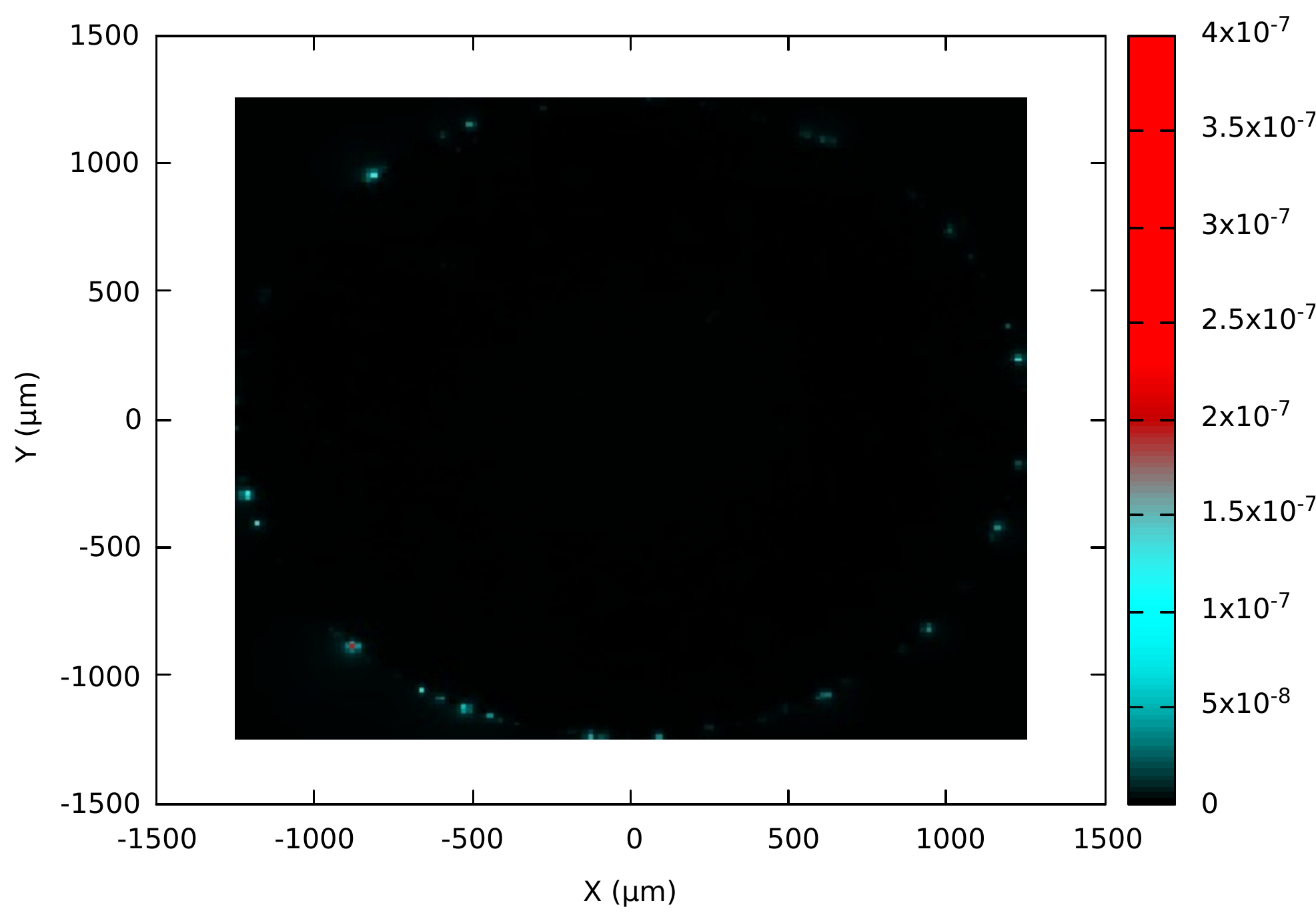}
\caption{A typical current map of sample C for a particular realization of the random distribution.
  Emission is mostly from a few spots on the cathode boundary due to the high density of emitters.
  Note that the cathode is circular in shape.}
\label{fig:map_C}
\end{center}
\end{figure}

A similar search for $R_a$ in case of sample C consisting of 44100 nanocone emitters
yields $R_a = 9.1$nm. Results for three realizations of emitter positions are shown
in Fig.~\ref{fig:sampleC}. It may be noted that there is a slight change in current
depending on the particular realization of random positions. The current map
from the LAFE shows that while sample B has substantial contribution to the current
form the interior of the LAFE (see Fig.~\ref{fig:map_B}), sample C, which has a much larger density, has large
contributions to the total LAFE current from a few isolated emitters on the boundary of
the circular patch and very little from the interior (see Fig.~\ref{fig:map_C}).
Since the number of emitters on the boundary is
statistically less, the LAFE current depends (albeit weakly) on each realization of emitter positions.

\begin{figure}[hbt]
  \begin{center}
    \vskip -0.5cm
\hspace*{-.50cm}\includegraphics[scale=0.33,angle=0]{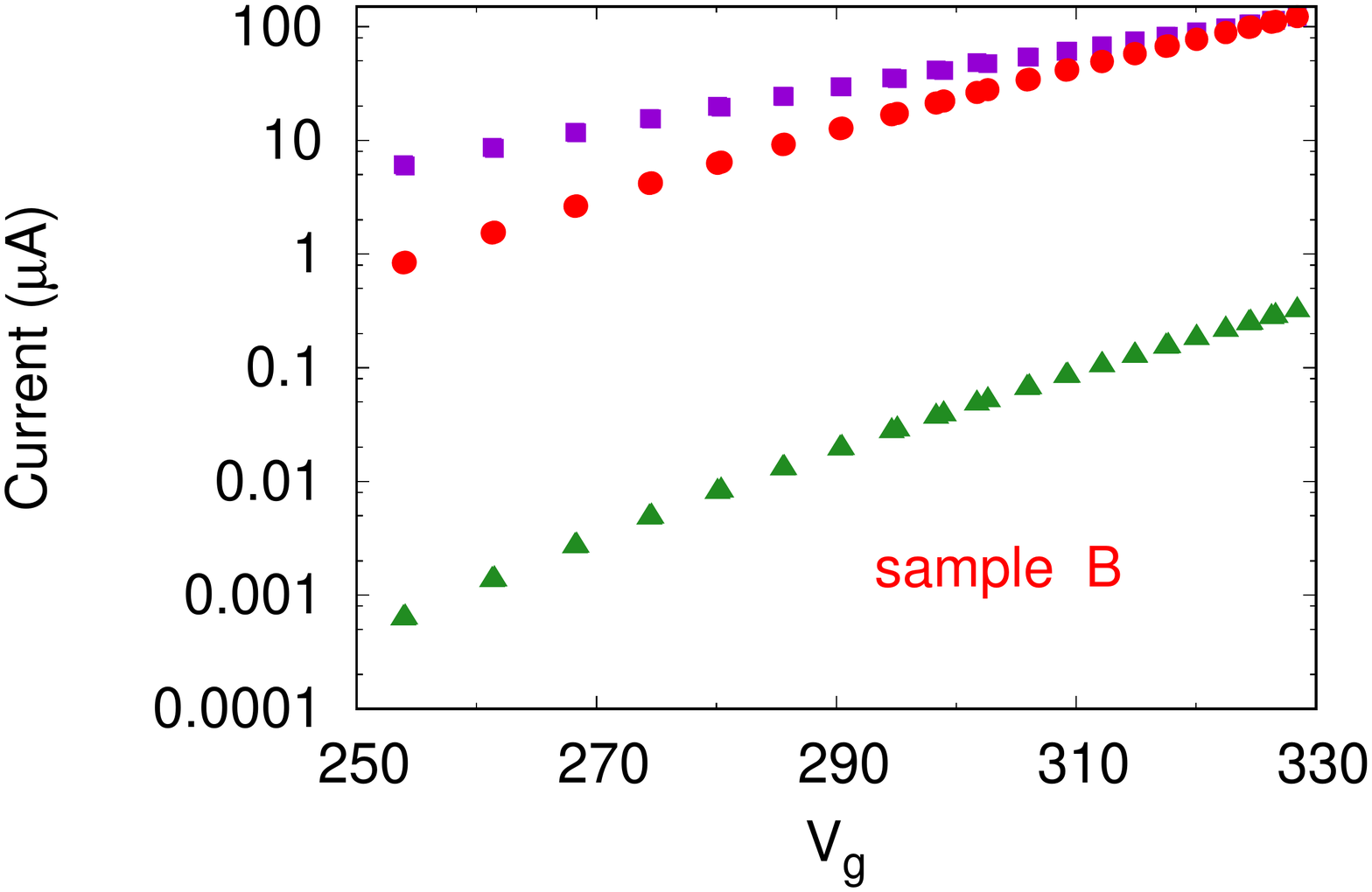}
\vskip -0.9 cm
\caption{The current from sample B. The solid squares are the experimental values
  while the circles represent the result of computation using the hybrid model with
  the anode at $D = 50\mu$m. The triangles are the result of the hybrid model with
  the anode far away.  }
\label{fig:sampleB_noanode}
\end{center}
\end{figure}

The hybrid model thus reasonably reproduces the $\text{I-V}$ curve for two samples with
widely different emitter-densities for values of apex radius of curvature $R_a$ that
are close to each other. The role of anode-proximity in field enhancement of a LAFE emitter can
be appreciated by comparing the results with a distant anode (i.e. by ignoring $\alpha_A$ and
$\alpha_{SA}$ in Eq.~(\ref{eq:gamSA}). Fig.~\ref{fig:sampleB_noanode} shows a
comparison of the experimental result for sample B (solid squares) with the results of the hybrid-model
when the anode is placed at $D = 50\mu$m from cathode (solid circle) and
for the anode far away (solid triangle). Clearly, ignoring the anode contribution to the
local field, would further lower
the apex radius of curvature ($R_a$) required to approximate the experimental result.
\vskip -0.1in
$\;$

\section{Discussions and Conclusions}

We have developed the hybrid model for dealing with a LAFE consisting of axially
symmetric emitters with smooth end-caps. The model allows the anode to be
placed even in close proximity.
It holds equally for ordered as well as random placement of emitters.
The model was tested for a LAFE ordered on an infinite square lattice and
was found to predict the apex field
enhancement factor with reasonable accuracy. It was also subjected
to experimental validation for a random LAFE.

In the absence of a definitive range of the apex radius of curvature ($R_a$) in the
experiment, an indirect approach towards validation had to be adopted.
Assuming a smooth emitter end-cap, it was demonstrated that the difference in $\text{I-V}$
characteristics of sample B and C can be explained largely on the basis of the difference
in emitter densities keeping individual emitters almost identical as in the experiment.
The hybrid model required that the apex radius of curvature of individual emitters differ mildly
in the two samples with sample C having a slightly higher value of $R_a$. We believe $R_a \simeq 9$nm
to be the highest value necessary to explain the experimental results if the end-cap is smooth.

As a precautionary note, taking a small representative sample
of emitters, rather than  the full LAFE, can grossly misrepresent the shielding and
anode-proximity effects since the boundary
gets a larger weight. Our simulations using the hybrid model  but smaller sample sizes having
the same density, show that optimum values of $R_a$ necessary to mimic the experimental $\text{I-V}$
curve can be much larger, depending on the size of the representative sample.
Such an approach is obviously flawed and conclusions based on it would be erroneous. In fact, the value of $R_a$
necessary to explain the experimental $\text{I-V}$ curve decreases as the sample size becomes
larger.

Finally, it is worth noting that the a perfect fit to the experimental data at all voltages $V_g$
was not possible by varying just the apex radius of curvature $R_a$. This is not a shortcoming
of the hybrid model and is indeed to be expected while dealing with emitters with apex radius
of curvature in the nanometer range. 
A previous study\cite{db_curvature} using the curvature corrected field emission current density, Eq.~(\ref{eq:FNC0}),
has shown that it works best at higher $V_g$ and $R_a$. The optimum value of $R_a$ was thus
chosen to reflect this fundamental property - higher values of $V_g$ having a better fit.
These findings underscore the need to introduce further corrections to the tunneling and image
potential\cite{db_imagepot,db_tunnelpot} and eventually to the current density.

In conclusion, we have demonstrated that the hybrid model, 
based on contemporary field emission theory with its recent advances
on shielding and anode-proximity, is fairly accurate in predicting local fields and
consistently explains experimental results on field emission from
a random distribution of gold nanocones. \\

{\it Acknowledgements}~\textemdash~ The author thanks Johannes Bieker for sharing the data reported
in Ref. [\onlinecite{bieker2018}] and SEM images of the nanocones. The author acknowledges several
discussions with Raghwendra Kumar and Rashbihari Rudra and helpful suggestions from  anonymous referees.

\section{References} 


\begin{thebibliography}{99}
\bibitem{teo}  K.~B.~K.~Teo, E.~Minoux, L.~Hudanski, F.~Peauger, J.~P.~Schnell, L.~Gangloff, P.~Legagneux, D.~Dieumegard, G.~A.~J.~Amaratunga and W.~I.~Milne, Nature 437, 968 (2005).
\bibitem{dams2012} F. Dams, A. Navitski, C. Prommesberger, P. Serbun, C. Langer, G.
  Muller, and R. Schreiner, IEEE Trans. Electron Devices 59, 2832 (2012).
\bibitem{wilfert2012} S.~Wilfert and C.~Edelmann, Vacuum 86, 556 (2012).
\bibitem{li2015} Y.~Li, Y.~Sun and J.~T.~W.~Yeow, Nanotechnology 26, 242001 (2015).  
\bibitem{hong2018} J.~H.~Hong, J.~S.~Kang and K.~C.~Par, J.~Vac.~Sci.~Tech. B, 36, 02C109 (2018).
\bibitem{sheshin2019} E.~P.~Sheshin, A.~Yu.~Kolodyazhnyj, N.~N.~Chadaev, A.~O.~Getman, M.~I.~Danilkin and
  D.~I. Ozol, J.~Vac.~Sci.~Tech. B, 37, 031213 (2019).
\bibitem{ohkawa2019} Y.~Ohkawa, T.~Okumura, K.~Iki, H.~Okamoto and S.~Kawamoto, J.~Vac.~Sci.~Tech. B, 37, 022203 (2019).
\bibitem{spindt76} C.~A.~Spindt, I.~Brodie, L.~Humphrey and E.~R.~Westerberg, J. Appl. Phys. 47, 5248 (1976).
\bibitem{spindt91} C.~A.~Spindt, C.~E.~Holland, A.~Rosengreen and I.~Brodie, IEEE Trans. on Electron Devices, 38, 2355 (1991).
\bibitem{whaley2009} D.~R.~Whaley, R.~Duggal, C.~M.~Armstrong, C.~L.~Bellew, C.~E.~Holland and C.~A.~Spindt, IEEE Trans. Electron Devices 56, 896 (2009).
\bibitem{helfenstein} P. Helfenstein, V. A. Guzenko, H. W. Fink, and S. Tsujino, J. Appl. Phys. 113, 043306 (2013).
\bibitem{read_bowring} F.~H.~Read and N.~J.~Bowring, Nucl. Instrum. Methods Phys. Res. A 519, 305 (2004).
\bibitem{bieker2018} J.~Bieker, F.~Roustaie, H.~F.~Schlaak, C.~Langer, R.~Schreiner, M.~Lotz and S.~Wilfert,
  J.~Vac.~Sci.~Tech. B 36, 02C105 (2018).
\bibitem{FN} R.~H.~Fowler and L.~Nordheim, Proc. R. Soc. A 119, 173 (1928).
\bibitem{MG} E.~L.~Murphy and R.~H.~Good, Phys. Rev. 102, 1464 (1956).
\bibitem{forbes_deane} R.~G.~Forbes and J.~H.~B.~Deane, Proc. Roy. Soc. A 463, 2907 (2007).
\bibitem{jensen_ency} K.~L.~Jensen, {\it Field emission - fundamental theory to usage}, Wiley Encycl. Electr. Electron. Eng. (2014).
\bibitem{kyritsakis2015} A.~Kyritsakis and J.~P.~Xanthakis, Proc. R. Soc. London, A471, 20140811 (2015).
\bibitem{db_parabolic} D.~Biswas, Phys. Plasmas 25, 043105 (2018).
\bibitem{db_curvature} D.~Biswas and R.~Ramachandran, J. Vac. Sci. Technol. B 37, 021801 (2019).
\bibitem{dist_known} The distribution of apex field enhancement factors due to shielding alone is known. See
  [\onlinecite{db_rudra}].
\bibitem{db_rudra} D.~Biswas and R.~Rudra, Physics of Plasmas 25, 083105 (2018).
\bibitem{levine95} J.~D.~Levine, J. Vac. Sci. Technol. B 13, 553 (1995).
\bibitem{bieker2019} J.~Bieker, R.~G.~Forbes, S.~Wilfert and H.~F.~Schlaak, IEEE Journal of the Electron Devices Society 7, 997 (2019).
\bibitem{deassiss2020} T.~A.~de Assiss, F.~F.~Dall'Agnol and M.~Cahay, Appl. Phys. Lett. 116, 203103 (2020).
\bibitem{assis2019} T.~A.~De~Assis and F.~F.~Dall’Agnol, J. Vac. Sci. Technol. B 37, 022902 (2019). 
\bibitem{db_fef} D.~Biswas, Phys. Plasmas 25, 043113 (2018).
\bibitem{db_anodeprox} D.~Biswas, Physics of Plasmas, 26, 073106 (2019).
\bibitem{rr_db_2019} R.~Rudra and  D.~Biswas, AIP Advances, 9, 125207 (2019).
\bibitem{db_rr_2020} D.~Biswas and R.~Rudra, J. Vac. Sci. Technol. B, 38, 023207 (2020).
\bibitem{db2016} D.~Biswas, G.~Singh and R.~Kumar, J.~App.~Phys 120, 124307 (2016).
\bibitem{mesa} E.~Mesa, E.~Dubado-Fuentes, and J.~J.~Saenz, J. Appl. Phys. 79, 39 (1996).
\bibitem{pogo2009} E.~G.~Pogorelov, A.~I.~Zhbanov, and Y.-C.~Chang, Ultramicroscopy 109, 373 (2009).
\bibitem{harris15} J.~R.~Harris, K.~L.~Jensen, D.~A.~Shiffler, and J.~J.~Petillo, Appl. Phys. Lett. 106, 201603 (2015).
\bibitem{harris16} J.~R.~Harris, K.~L.~Jensen, W.~Tang and D.~A.~Schiffler, J. Vac. Sci. Technol. B 34, 041215 (2016).
\bibitem{alp1alp2} The coefficients $\alpha_1$ and $\alpha_2$ are known for special shapes for which the
  problem is exactly solvable such as the hemiellipsoid. In the framework of the line charge model, the
  shape is determined by the zero-potential contour generated by the line charge density and the
  macroscopic field. Thus the geometry is encoded within the line charge density.
  When the problem is solvable, the equivalent line charge density can be determined
  from the surface charge density $\sigma = \epsilon_0 \vec{E}.\hat{n}$. See Ref.~[\onlinecite{db2016}] for an illustration.
  A general expression for $\alpha_1$ and $\alpha_2$ in terms of the line charge density can be found in
  [\onlinecite{db_fef}].
\bibitem{edgcombe2002} C.~J.~Edgcombe, and U.~Valdr\`{e}, Philosophical Magazine B 82, 987 (2002).
\bibitem{forbes2003} R.~G.~Forbes, C.~J.~Edgcombe, and U.~Valdr\`{e}, Ultramicroscopy 95, 57 (2003).
\bibitem{shreya_db_2019} S.~Sarkar and D.~Biswas, J. Vac. Sci. Technol. B37, 062203 (2019).
\bibitem{wang} X.~Q.~Wang, M.~Wang, P.~M.~He, Y.~B.~Xu, and Z.~H.~Li, J. Appl. Phys. 96,
6752 (2004).
\bibitem{smith} R.~C.~Smith, D.~C.~Cox, and S.~R.~P.~Silva, Appl. Phys. Lett. 87, 103112 (2005).

\bibitem{error1} The error increases as $c$ is reduced since the zero potential
  contour generated by an isolated line charge density, does not retain the shape of the emitter. 
\bibitem{db_ultram} D.~Biswas, G.~Singh, S.~G.~Sarkar and R.~Kumar, Ultramicroscopy 185, 1 (2018).
\bibitem{cosine} D.~Biswas, G.~Singh and R.~Ramachandran, Physica E 109, 179 (2019).
\bibitem{schott23} W. Schottky, Z. Phys. 14, 63 (1923).
\bibitem{stern} T.~E.~Stern, B.~S.~Gossling and R.~H.~Fowler, Proc. R. Soc. Lond. A 124, 699 (1929).
\bibitem{db_schottky} D.~Biswas, J. Vac. Sci. Technol. B 38, 023208 (2020).
\bibitem{db_gated} D.~Biswas and R.~Kumar, J. Vac. Sci. Technol. B 37, 040603 (2019).
\bibitem{db_imagepot} D.~Biswas and R.~Ramachandran, Phys. Plasmas, 24, 073107 (2017).
\bibitem{db_tunnelpot} D.~Biswas, R.~Ramachandran and G.~Singh, Phys. Plasmas, 25, 013113 (2018).

  
\end{thebibliography}
\end{document}